\documentclass[%
 preprint,
superscriptaddress,
 amsmath,amssymb,
 aps,
 prapplied
]{revtex4-2}

\usepackage{graphicx}
\usepackage{dcolumn}
\usepackage{bm}
\usepackage{hyperref}
\usepackage{xcolor}

\begin{document}

\preprint{APS/123-QED}

\title{Towards compact high-frequency nonreciprocal devices using nanoplasma-switched time-varying metasurfaces}

\author{Mikhail Sidorenko}
\email{mikhail.sidorenko@aalto.fi}
\affiliation{Department of Electronics and Nanoengineering, Aalto University, Espoo, Finland}%

\author{Jin Zhang}
\affiliation{Department of Electronics and Nanoengineering, Aalto University, Espoo, Finland}%

\author{Xuchen Wang}
\affiliation{Qingdao Innovation and Development Base
Harbin Engineering University
266400, Qingdao, China}%

\author{Zhipei Sun}
\affiliation{Department of Electronics and Nanoengineering, Aalto University, Espoo, Finland}%

\author{Sergei Tretyakov}
\affiliation{Department of Electronics and Nanoengineering, Aalto University, Espoo, Finland}%

\author{Viktar Asadchy}
\email{viktar.asadchy@aalto.fi}
\affiliation{Department of Electronics and Nanoengineering, Aalto University, Espoo, Finland}%

\date{\today}

\begin{abstract}
Time-modulated systems have received growing interest in recent years. They allow us to tailor effects, such as frequency conversion, single-direction propagation, etc. For the microwave band, semiconductor elements, such as varactors, are usually used as time-modulated elements but their modulation frequency has been limited to the few-gigahertz range. Recent advances in nanoplasma switches, i.e., two-state electronic switches based on a gas discharge in a nanometer-scale gap, provide a new potential for developing time-modulated systems with high operating frequencies. Here, we develop an analytical framework based on the time-Floquet method for the design of nonreciprocal time-modulated devices based on two-state time-modulated elements, for instance, nanoplasma-based switches. A practical example of a microwave isolator operating at 100~GHz frequency is developed and studied both analytically and using full-wave simulations. A potential realization in a parallel-plate waveguide is also simulated numerically.
\end{abstract}

\maketitle

\section{Introduction}

Electromagnetic Lorentz reciprocity is a fundamental symmetry of Maxwell’s equations. It states that in linear, time-invariant, and passive media governed by time-reversal-symmetric equations, the transmission characteristics between two points remain unchanged when the positions of the source and receiver are interchanged. This principle defines the behavior of most conventional antennas, waveguides, and scattering systems. Violation of Lorentz reciprocity arises when these conditions are broken, for example, by magnetic field biasing, temporal modulation, or nonlinear interactions, enabling direction-dependent wave propagation~\cite{Asadchy2020TutorialNonreciprocity,Caloz2018ElectromagneticNonreciprocity}. Such nonreciprocal electromagnetic responses have enabled a range of applications, including isolators and circulators, unidirectional antennas and metasurfaces, and topological photonic devices.

Microwave isolators form a particular class of nonreciprocal devices. They are two-port devices designed to provide unidirectional power transmission at the prescribed frequency band. Microwave isolators are widely used in many practical applications, such as mobile and satellite communications and radar systems. A typical commercial isolator is based on ferrite materials and provides about $20 \,\mathrm{dB}$ isolation with an insertion loss often less than $1 \, \mathrm{dB}$ \cite{ref4}. Ferrite materials provide nonreciprocal behavior under an external biasing magnetic field. However, in some applications, magnetic materials are undesirable for a wide class of reasons. For such applications, magnetless isolators have been developed for several decades. They can be based on active elements, semiconductor frequency mixers \cite{IEEE1}, nonlinear effects \cite{Nat3}, time-modulated resonators \cite{Nat2} or topological effects \cite{Nat1}. 

In all nonreciprocal systems based on time modulation, the choice of the time-modulated element is critical. For microwave implementations, PIN diodes, varactor diodes, and MOSFET switches are commonly employed~\cite{elnaggar2020modeling,wang2020nonreciprocity,reiskarimian2016magnetic,dinc2017synchronized}. However, achieving large modulation depths of capacitance or conductance at high modulation rates with available components remains highly challenging. Varactor-based nonreciprocal filters and circulators typically operate with modulation frequencies in the $10^7\text{--}10^8~{\rm Hz}$ range~\cite{chaudhary2022frequency,chaudhary2024magnetless}, whereas integrated CMOS transistor-switch implementations can reach modulation frequencies in the $10^8\text{--}10^9~{\rm Hz}$ range~\cite{dinc2017synchronized}. 
These limitations not only restrict performance well below the frequency regimes targeted by many next-generation platforms, but also significantly hinder the scalability of time-modulated isolators to higher-frequency bands.
Nonreciprocal behavior can also be achieved with modulation frequencies several orders of magnitude lower than the signal frequency by employing traveling-wave spatiotemporal modulation. However, such demonstrations have required extremely long modulation sections, sometimes extending over thousands of wavelengths~\cite{ref1}. While this approach is feasible in optical platforms, it is generally impractical at microwave frequencies due to prohibitive physical dimensions. 

On the other hand, recent progress in the plasma community, particularly the emergence of nanoscale plasmas (nanoplasmas), has revealed a promising route toward high-frequency time-varying microwave systems. Enabled by advances in nanofabrication, nanoplasma discharges can be driven on picosecond timescales and potentially faster \cite{ref2,ref10,ref11,ref12,ref13,ref14}. A nanoplasma discharge is a plasma discharge formed in an ultrashort gap, where the electrode separation is on the order of a few micrometers or even hundreds of nanometers. In such a gap, the discharge ignition process is mainly driven by field emission effect electrons. Due to the very small gap size, the electric field strength is high, and electrons can leave the cathode under normal conditions without additional heating. This process is called \emph{electron field emission} and is described by the Fowler-Nordheim equation \cite{FoNo}. Although the maximal current provided by this effect is extremely small under realistic conditions, the emitted electrons start avalanche ionization, and the nanoplasma discharge is ignited at a time scale of picoseconds or less. Thus, a nanoplasma discharge can be regarded as an ultrafast electronic switch~\cite{ref2} with two distinct states, ON and OFF, thereby providing a time-modulated conductance that can potentially reach modulation speeds up to the few-THz range. To date, reported applications of nanoplasma have been largely focused on ultrafast switching for THz pulse generation \cite{ref2,ref13} and on creating time interfaces in wire media with strong spatial dispersion \cite{our1,our2,our3}, while its potential as an enabling mechanism for nonreciprocal devices has not yet been explored.

In this article, we propose the use of nanoplasma for creating high-frequency magnetless nonreciprocal components with a compact overall size. We show both theoretically and numerically a design of an isolator operating at a frequency of 100~GHz formed by cascading \textit{only two} time-modulated metasurfaces whose equivalent sheet conductances are driven at the same modulation frequency but with a controlled relative phase. Each metasurface comprises a periodic array of metallic strips loaded by nanoplasma discharge gaps, enabling a \textit{stepwise} (binary) temporal modulation of the sheet impedance. This switching modality leads to an analytical treatment and an equivalent-circuit description that strongly differ from prior harmonic time-modulated nonreciprocal metasurface implementations relying on semiconductor tuning elements. We further propose a compact realization of the isolator in a parallel-plate waveguide and verify it using full-wave simulations, demonstrating an isolation of 23 dB with an insertion loss of 2 dB for a total device thickness below 1.8 wavelengths. Overall, the proposed nanoplasma-based approach provides a new pathway toward nonreciprocal components operating at millimeter-wave frequencies and beyond.

\section{Theory}

\subsection{Static metasurface: grid of strips}\label{infinite-grid-of-wires}

\begin{figure}[b]
\begin{center}
\includegraphics[width=0.49\textwidth]{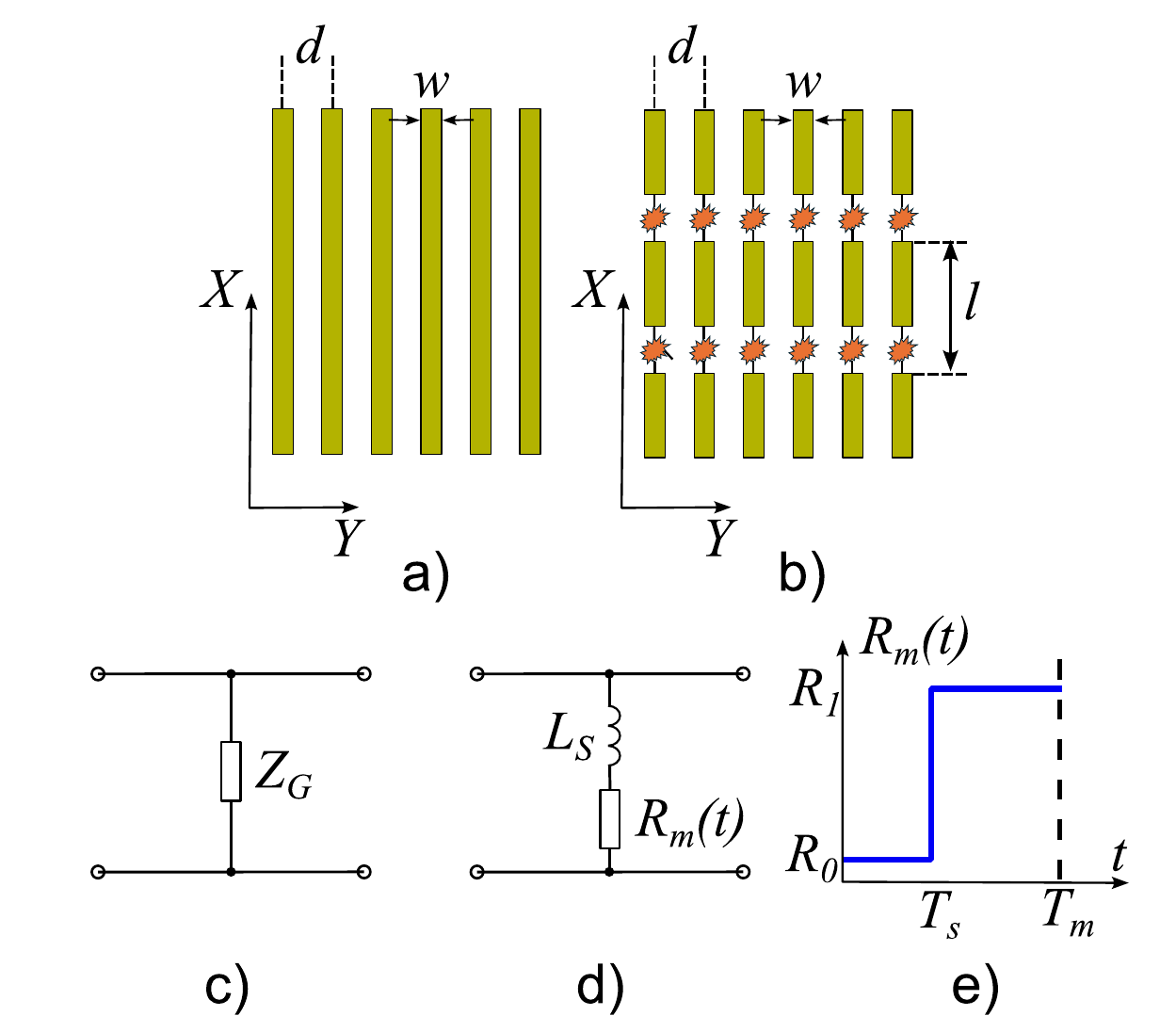}
\caption{\label{fig:Fig1} a) A schematic view of a strip grid; b) a schematic view of a strip grid switched by nanoplasma discharges; c) an equivalent circuit scheme of a strip grid; d) an equivalent circuit scheme of a time-switched strip grid; e) a time modulation function of $R_{\mathrm m}(t)$.}
\end{center}
\end{figure}

Let us consider an infinite flat array of periodic parallel strips of width $w$ in the plane $z = \mathrm{const}$ (see Fig. \ref{fig:Fig1}a), with the period $d \gg w$, and the thickness of the strips much smaller than $w$. For an $x$-polarized plane wave with the circular frequency $\omega_0$ and the corresponding wavelength $\lambda = 2\pi c/\omega_0 \gg d$ propagating in the $z$ direction, an analytic formula for the reflection and transmission coefficients is obtained with the help of an equivalent circuit model presented in Fig. \ref{fig:Fig1}c. For ideally conducting strips, the equivalent surface impedance of the grid is defined by its effective inductance and impedances of the periodically inserted loads~\cite{ref6}:
\begin{equation}\label{L_strips}
Z_G= Z_Ld + j\omega_0 L_\mathrm{s} = Z_Ld + j\omega_0\frac{d\mu_0}{2\pi}\log{\frac{1}{\displaystyle \sin{\frac{\pi w}{2d}\ \ }},}
\end{equation} 
where $Z_L$ is the load impedance per unit length, and  $j$ is the imaginary unit.

The $ABCD$ transfer matrix of an infinite wire grid in the free space has the form
\begin{equation} 
{\overline{\overline{T}}}_G = \begin{pmatrix}
1 & 0 \\
Y_G & 1 \\
\end{pmatrix},\ \ \ Y_G = Z_G^{- 1}.
\end{equation} 
The S-parameters then can be calculated (see, for instance, \cite{ref7}):
\begin{equation} 
S_{11} = - \frac{Z_G^{-1}Z_0}{2 + Z_G^{-1}Z_0},\ \ S_{12} = \frac{2}{2 + Z_G^{-1} Z_0},
\end{equation} 
where $Z_0$ is the free-space impedance.

\subsection{Time-modulated metasurface: a grid with switchable nanogaps}\label{time-modulated-grid-with-switches}

To obtain magnetless nonreciprocal behavior, we consider the time-modulated case. The grid parameters are changed periodically with the temporal period \(T_{\rm m}=2\pi/\omega_{\rm m}\). Each strip is cut into pieces of equal length \(l\), with nanoscale gaps between adjacent pieces. An external biasing voltage is applied to the gaps, so that a nanoplasma discharge can be ignited in each gap. Once ignited, the nanoplasma discharges effectively transform the broken
strip into a continuous conducting strip. After the biasing voltage is removed, the nanoplasma discharges decay, and the strip returns to its initial broken state. This grid is schematically depicted in Fig.~\ref{fig:Fig1}b. According to Ref.~\cite{ref2}, the ignition transient of a nanoplasma discharge takes only several picoseconds, which is shorter than the period of the incident wave \(T_0=2\pi/\omega_0\), therefore, the switching can be considered
instantaneous on the RF time scale. Although experimental measurements of the nanoplasma discharge impedance at high frequencies are not yet available, the numerical and experimental results reported for millimeter-scale plasma gaps \cite{plamsa_study_1,plasma_study_2,plasma_study_3} suggest that, at
frequencies on the order of \(100~{\rm GHz}\), the nanoplasma gap impedance is predominantly resistive, while its inductive and capacitive contributions can be neglected.

To develop an analytical model, first, we need to choose the physical model of the time-modulated nanoplasma discharge. Since we consider two states of nanoscale gaps (a conductor when a nanoplasma discharge is ignited and an isolator when there is no discharge), the gap acts like a two-state switch. A model of a time-modulated switch was proposed in \cite{switches1}.  Since, as we stated above, we assume the nanoplasma discharge to have a  purely resistive impedance, we propose to use a simplified model with a time-modulated resistor that takes only two values -- $R_0$ and $R_1$:
\begin{equation} 
R_{\rm m}(t) = \left\{ 
\begin{array}{ll}
R_{0}, & 0 \leq t < T_{\rm s} \\
R_{1}, & T_{\rm s} \leq t < T_{\rm m} 
\end{array},
\quad R_{\rm m}\left( t + T_{\rm m} \right) = R_{\rm m}(t), \right. 
\end{equation} 
and $T_{\rm s} < T_{\rm m}$ governs the moment of switching within the modulation period $T_{\mathrm m}$. For convenience, we introduce a new parameter
\begin{equation}\label{xi}
\xi = \frac{T_{\rm s}}{T_{\rm m}}.    
\end{equation}
The time plot of such a function is schematically presented in Fig.~\ref{fig:Fig1}e. When the switch is in the ON stage, we consider it as a resistor with a purely resistive impedance $R_0$, which is very small and could be considered equal to zero. 
When the switch is in the OFF stage, the resistance of the gap is infinite. However, since we are going to use this model for a numerical optimization, we cannot set the resistance equal to infinity. Instead, we set it equal to an arbitrary very large value. To ensure that the chosen value is sufficient from the point of view of a numerical evaluation, we calculate the value of $|S_{11}|$ of a static grid with the chosen gap resistance numerically, and then calculate this value again, but for the gap resistance 10 times larger. If the relative change of $|S_{11}|$ is below $0.01\%$, we use this large resistance as a substitution for the infinite resistance in our numerical calculations. 

Now we have a final equivalent circuit for an approximate model of a two-state strip grid, which is presented in Fig.~\ref{fig:Fig1}(d). Since this circuit contains both resistive and reactive elements, we use the Fourier series for each time-dependent value to present the relation between the current and the voltage.
To obtain an analytical result, we apply the time-Floquet analysis \cite{ref8}, \cite{ref9}. The periodic function $R_{m}(t)$ can be easily presented in the form of a series
\begin{equation} 
R_{\rm m}(t) = \sum_{m = - \infty}^{\infty}{r_{m}\exp{\left( jm\omega_{\rm m} t \right).}} \label{eq: R}
\end{equation} 
For the chosen stepwise function $R_{\rm m}(t)$ straightforward calculation shows that the coefficients have the form

\begin{equation}\label{eq_rm}
\begin{array}{rcl}
r_m & = & \left\{ \begin{array}{ll}
     \displaystyle \frac{j}{T_{\rm m} m\omega_{\rm m}} \left\{ R_0\left\lbrack \exp{\left( - jm\omega_{\rm m} T_{\rm s} \right) - 1} \right\rbrack  \right. & \\ \\
      + \left. R_1\left\lbrack \exp\left( - jm\omega_{\rm m} T_{\rm m} \right) - \exp{( - jm\omega_{\rm m} T_{\rm s} )} \right\rbrack \right\}, & m\neq 0, \\ \\
      \displaystyle \frac{1}{T_{\rm m}} \left[ R_0 T_s + R_1(T_{\mathrm m} - T_s) \right], & m=0. \end{array}  \right.
\end{array}
\end{equation} 

The voltage $V(t)$ and the current $I(t)$ across the time-varying resistor are also expanded into the series:
\begin{equation} 
\begin{array}{ll}
I(t) = \displaystyle \sum_{n = - \infty}^{\infty}{i_n \exp(jn\omega_{\rm m}t)} \\  \displaystyle V(t) = \sum_{n = - \infty}^{\infty}{v_n \exp(jn\omega_{\rm m} t)}. 
\end{array} \label{eq: VI}
\end{equation} 
By substituting Eqs.~(\ref{eq: R}) and~(\ref{eq: VI}) into 
Ohm's law $V(t)=I(t)R(t)$, we obtain the following equation:
\begin{equation} \label{eq_full}
\begin{array}{l}
\displaystyle \sum_{n = - \infty}^{\infty}{v_{n} \exp(jn\omega_{\rm m}t)} \\ 
\displaystyle = \sum_{n = - \infty}^{\infty\ }{\sum_{m = - \infty}^{\infty}{r_mi_{n}\exp\left( j(n + m)\omega_{\rm m} t \right)}}.
\end{array}
\end{equation} 
By replacing $n$ with $n - m$ in the right-hand side of (\ref{eq_full}) we obtain a set of equations for each $n$:
\begin{equation}\label{eq_set1}
\sum_{m = - \infty}^{\infty}r_m i_{n - m} = v_n, \qquad n=-\infty \, \ldots +\infty.
\end{equation} 

Next, we cut the number of harmonics under consideration from an
infinite number to $2N + 1$, since the coefficients (\ref{eq_rm}) decay as $n \to\infty $. Then the infinite set of equations (\ref{eq_set1}) is represented in matrix form:
\begin{equation} \label{eq_matrix1}
\begin{array}{l}
{\overline{\overline{Z}}}_{R} \cdot \overline{i} = \overline{v}, \\{\overline{\overline{Z}}}_{R} = \left(
\begin{array}{ccccccc}
r_0 & & r_{- 1} & \ldots & r_{-2N+1} & & r_{-2N}\\
r_1 & & r_0 & & r_{-2N+2} & & r_{-2N+1} \\
&\vdots  & & \ddots & & \vdots & \\
r_{2N-1} & & r_{2N-2} & & r_0 & & r_{-1} \\
r_{2N} & & r_{2N-1} & \ldots & r_1 & & r_0
\end{array} \right),
\end{array}
\end{equation} 
where the matrix ${\overline{\overline{Z}}}_{R}$ is built from the Fourier series coefficients of $R_{\rm m}(t)$.

For the elements of the equivalent circuit that do not depend on time, the matrix for the time-Floquet method is a diagonal matrix. For the equivalent inductance the matrix has the form \cite{ref9}
\begin{equation}\label{eq_martix_Zl}
{\overline{\overline{Z}}}_{L} = jL_{s}\left(
\begin{array}{ccc}
\omega_0-N\omega_{\rm m} & \ldots & 0 \\
\vdots & \ddots & \vdots \\
0 & \ldots & \omega_0 + N\omega_{\rm m}
\end{array}
\right),
\end{equation} 
where the grid inductance $L_s$ is defined by (\ref{L_strips}).
By combining the matrices ${\overline{\overline{Z}}}_{R}$ and
${\overline{\overline{Z}}}_{L}$ we obtain the
$2(2N + 1) \times 2(2N + 1)$ transfer matrix
${\overline{\overline{T}}}_{G}$ for the time-modulated grid in the
form
\begin{equation} 
{\overline{\overline{T}}}_{G} = \left(\begin{array}{cc}
\overline{\overline{I}} & \overline{\overline{0}} \\
{\overline{\overline{Y}}}_{G} & \overline{\overline{I}} \\
\end{array}\right) ,\quad {\ \ \ \ \ \ \overline{\overline{Y}}}_{G} = \left( {\overline{\overline{Z}}}_{R} + {\overline{\overline{Z}}}_{L} \right)^{- 1},
\end{equation} 
where $\overline{\overline{I}}$ is the ($2N + 1) \times (2N + 1)$
identity matrix. This matrix sets the relation between the harmonics of the input and output voltages and currents of the equivalent circuit.

\subsection{Multi-layered metasurface}\label{multiple-layers}

Let us now consider the case when the strip grid is built on a 
dielectric substrate. The formula for the transfer matrix of a dielectric slab of thickness $h$ and relative dielectric permittivity
$\varepsilon_{d}$ in free space can be found e.g. in \cite{ref7} (extended to the time-Floquet analysis in \cite{ref9}):
\begin{equation} 
{\overline{\overline{T}}}_d = {1\over 2}\left(
\begin{array}{cc}
\displaystyle  {\overline{\overline{P}}}_d + {\overline{\overline{P}}}_d^{- 1} & \displaystyle  - {\overline{\overline{Z}}}_d\left( {\overline{\overline{P}}}_d - {\overline{\overline{P}}}_d^{- 1} \right) \\
\displaystyle  - {\overline{\overline{Z}}}_d^{- 1}\left( {\overline{\overline{P}}}_d - {\overline{\overline{P}}}_d^{-1} \right) & \displaystyle   {\overline{\overline{P}}}_d + {\overline{\overline{P}}}_d^{- 1} 
\end{array} 
\right),
\end{equation} 
where the matrix ${\overline{\overline{P}}}_{d}$ is a
($2N + 1) \times (2N + 1)$ diagonal matrix with the elements on the
main diagonal
\begin{equation} 
\begin{array}{l}
\displaystyle
P_d(n,n) = \exp\left( -j\frac{\omega_n}{c}\sqrt{\varepsilon_d}\, h \right), \\ \\[2mm]
\displaystyle
\omega_n = \omega_0+(n-N-1)\omega_{\mathrm{m}}, 
\quad n=1,\ldots,2N+1 .
\end{array}
\end{equation}
Here, $c$ is the speed of light, and the matrix ${\overline{\overline{Z}}}_{d}$ is a ($2N + 1) \times (2N + 1)$ diagonal matrix with the elements on the main diagonal equal to
\begin{equation} 
Z_d(n,n) = \frac{\mu_{0}c}{\sqrt{\varepsilon_d}},\quad n = 1\ldots 2N + 1.\ 
\end{equation} 

For the system comprised of $M$ layers, we multiply the transfer matrices. The direction of propagation defines the order in which the matrices are multiplied. The result gives two $2(2N + 1) \times 2(2N + 1)$ matrices for the forward and backward propagations, with the direction indicated by a superscript:
\begin{equation} \label{eq_matrices_tot}
{\overline{\overline{T}}}_\mathrm{tot}^\mathrm{f} = \prod_{p = M}^{1}{\overline{\overline{T}}}_{p},\quad {\overline{\overline{T}}}_\mathrm{tot}^\mathrm{b} = \prod_{p = 1}^{M}{{\overline{\overline{T}}}_{p},}
\end{equation} 
where ${\overline{\overline{T}}}_{p},\ \ p = 1\ldots M$ are the transfer matrices for each layer.

\subsection{Time-modulated multi-layered metasurface and harmonics generation}

Now let us relate the currents and voltages in the equivalent circuit to the electric and magnetic fields interacting with a multilayered metasurface. We consider an incident plane wave of frequency \(\omega_0\), whose electric field is polarized parallel to the strips. Let \(\omega_{\rm m}\) be the modulation frequency of the metasurface. Since the system is periodic in time, an incident wave at \(\omega_0\) can couple only to temporal Floquet harmonics with frequencies
\begin{equation}
    \omega_n=\omega_0+n\omega_{\rm m}, \qquad n=-N,\ldots,N .
\end{equation}
Accordingly, we represent the field amplitudes in the truncated Floquet basis as
\begin{equation}
    E(t)=\Re\left[
    \sum_{n=-N}^{N} e_n \exp\left(j\omega_n t\right)
    \right],
    \qquad
    e_n \equiv e(\omega_0+n\omega_{\rm m}) .
\label{eq:floquet_field_expansion}
\end{equation}
For a monochromatic incident wave, only the central Floquet channel is excited; therefore, with the normalization \(e_0=1\), the incident-amplitude vector is
\begin{equation}
    \overline{e}_{i}
    =
    (0,\ldots,0,1,0,\ldots,0)^{\rm T}.
\label{eq:incident_floquet_vector}
\end{equation}
Let us also consider transmitted field $E_\mathrm{t}$ and reflected field $E_\mathrm{r}$. For a static non-modulated system, the vectors for transmitted and the reflected waves will also have only one non-zero coefficient each, corresponding to the frequency $\omega_0$. But in a time-modulated case, new harmonics arise:
\begin{equation} 
\begin{array}{rcl}
\overline{e}_t & = & \left( e_t\left( \omega_0 - N\omega_{\mathrm m} \right),\ \ldots,\ e_t\left( \omega_0 - \omega_{\mathrm m} \right), \right. \\ \\
& &\left. e_t\left( \omega_0 \right),e_t\left( \omega_0 + \omega_{\mathrm m} \right),\ \ldots, e_t(\omega_0 + N\omega_{\mathrm m}) \right),
\end{array}
\end{equation} 
\begin{equation} 
\begin{array}{rcl}
\overline{e}_r & = & \left( e_r\left( \omega_0 - N\omega_{\mathrm m} \right),\ \ldots,\ e_r\left( \omega_0 - \omega_{\mathrm m} \right), \right. \\ \\
& & \left. e_r\left( \omega_0 \right),e_r\left( \omega_0 + \omega_{\mathrm m} \right),\ \ldots,\ e_r(\omega_0 + N\omega_{\mathrm m}) \right).
\end{array}
\end{equation} 
Note that these formulas include both positive and negative frequencies. When the frequency of a certain harmonic is negative, it means that the corresponding complex-valued exponent has the complex conjugate phase. However, since only the real values of the electric field have physical sense, these harmonics will contribute to the output in the same way as the harmonics with positive frequencies.

To find the relations between the vectors $\overline{e}_i, \overline{e}_t$ and $\overline{e}_r$ we build $(2N+1) \times (2N+1)$ transmission and reflection matrices, following \cite{ref9}:
\begin{equation} 
\begin{array}{l}
\overline{\overline{T}} = 2\left( \overline{\overline{A}} + \overline{\overline{B}} \,{\overline{\overline{Y}}}_0 + {\overline{\overline{Z}}}_0\overline{\overline{C}} + {\overline{\overline{Z}}}_0\overline{\overline{D}}\,{\overline{\overline{Y}}}_0 \right)^{-1}, \\ 
\overline{\overline{R}} = \left( \overline{\overline{A}} + \overline{\overline{B}}\,{\overline{\overline{Y}}}_0 \right)\overline{\overline{T}} - \overline{\overline{I}}.
\end{array}
\end{equation} 
Here, ${\overline{\overline{Y}}}_{0} = {\overline{\overline{Z}}}_{0}^{- 1}$
is the admittance matrix of free space, $\overline{\overline{I}}$
is the ($2N + 1) \times (2N + 1)$ identity matrix, and the matrices
$\overline{\overline{A}},\overline{\overline{B}},\overline{\overline{C}},\overline{\overline{D}}$ are the ($2N + 1) \times (2N + 1)$ blocks that form the total transfer matrices (\ref{eq_matrices_tot}) for the forward and the backward incidence, respectively.

\section{Numerical optimization}
\subsection{Design of a time-modulated metasurface isolator}

An electromagnetic isolator is a nonreciprocal two-port device designed to allow wave transmission in one direction and strongly attenuate or block transmission in the opposite direction. In the ideal case, it is described by the following $S$-matrix:
\begin{equation}
    \mathrm{\mathbf{S}}_{\mathrm{Isolator}} \equiv \left( \begin{array}{cc}
     S_{11}    &  S_{12} \\
     S_{21}    &  S_{22}
    \end{array} \right) = \left( \begin{array}{cc}
     0    &  0 \\
     1    &  0
    \end{array} \right).
\end{equation}

Let the incident wave has the frequency $\omega_0$ and the metasurface is modulated at the frequency $\omega_{\mathrm m}$, $n\omega_{\mathrm m} \neq 2\omega_0$ for any integer $n$. This condition ensures that none of the harmonic frequencies $\omega_0 \pm n\omega_{\mathrm m}, $
$n = 1\ldots N$ coincides with $-\omega_0$.

We denote as ${\overline{e}}_{r}^\mathrm{f}$ the vector of the amplitudes of the harmonics of the forward reflected field, i.e., the reflection from the time-modulated structure when it is illuminated by a wave traveling from $z=-\infty$ to $z=+\infty$  and ${\overline{e}}_{t}^\mathrm{b}$ the vector of   harmonic amplitudes at backward transmission, i.e., the transmission of  harmonics when the time-modulated structure is illuminated by a wave traveling from $z=+\infty$ to $z=-\infty$ (see Fig.~\ref{fig:Fig2}c).

\subsubsection*{Design goals}
\begin{enumerate}
\def\labelenumi{\arabic{enumi}.}
\item
  For an incident wave with the frequency $\omega_0$ in the `forward' direction we want $e_t^\mathrm{f}\left( \omega_0 \right)$ to be as close to $1$ as  possible, which is equivalent to low insertion loss of the isolator. For a time-modulated system, the energy in the general case is not conserved even in the absence of losses, so this coefficient may be larger than 1.
\item
  For an incident wave with the frequency $\omega_0$ in the `forward' direction we want $e_r^\mathrm{f}\left( \omega_0 \right)$ to be as close to $0$ as possible, which means that no energy is reflected. Since the energy in this time-varying system is not generally conserved, we need to pose this goal in addition to goal 1.
\item
  For an incident wave with the frequency $\omega_0$ in the `backward' direction we want $e_t^\mathrm{b}\left( \omega_0 \right)$ to be as close to $0$ as possible, which means that no energy at the frequency $\omega_{0}$ can travel backward. In our case, it is not important if this energy is reflected or absorbed.
\item
  All other coefficients of the vectors ${\overline{e}}_t^\mathrm{f}$, ${\overline{e}}_r^\mathrm{f}$, ${\overline{e}}_t^\mathrm{b}$ may be arbitrary. All the  frequency side harmonics can be efficiently filtered out by any known passive band-pass filter with the center
  frequency $\omega_0.$ 
\end{enumerate}

\begin{figure*}
\begin{center}
\includegraphics[width=0.99\textwidth]{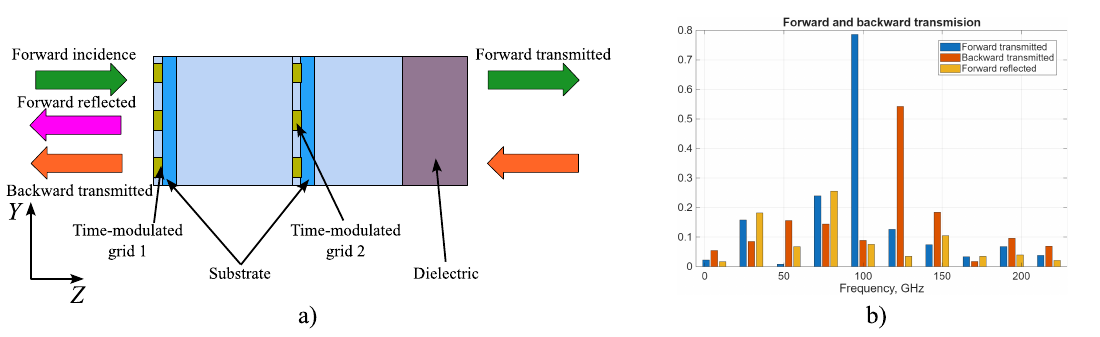}
\caption{\label{fig:Fig2} a) A schematic view of a multi-layered isolator device; b) the amplitude ratio for transmission and reflection for the multi-layered structure after MATLAB optimization, as predicted by the theory.}
\end{center}
\end{figure*}

We consider the following combination of the layers of different types in the order of increase of the $z$-coordinate (see 
Fig.~\ref{fig:Fig2}a): 1) a time-modulated strip grid; 2) a dielectric substrate (sapphire, $\varepsilon=9.3$ for a normal incidence on a c-plane wafer, losses are negligible); 3) an air gap; 4) a time-modulated strip grid; 5) a dielectric substrate (sapphire); 6) an air gap; 7) a dielectric slab (PTFE, $\varepsilon=2.1$, losses are negligible). Sapphire was chosen because this material can form a nearly perfect dielectric substrate for nanofabrication; PTFE was chosen because of its very low dielectric losses. Two time-modulated grids are the minimal quantity of time-modulated elements necessary to obtain nonreciprocal behavior~\cite{Asadchy2020TutorialNonreciprocity}. An additional dielectric slab strengthens the asymmetry of the system. For the sake of simplicity, we use the same modulation frequency $\omega_{\mathrm m}$ for both grids. 

We note here, that the formula (\ref{L_strips}) is, generally, valid only for grids in  free space or for a grid on an electrically thin dielectric substrate. A rigorous analysis of the impact of the finite-thickness slab on the grid impedance goes beyond the scope of the present article. However, for a substrate thickness close to $\lambda_d/2$, where $\lambda_d$ is the wavelength of the main harmonic inside the dielectric, the slab becomes effectively transparent. We have checked numerically that  formula (\ref{L_strips}) can be used in this case. For $\omega_0 = 2\pi \cdot 100\cdot10^9 \, \mathrm{Rad/s}$, a sapphire slab of thickness $0.5 \, \mathrm{mm}$ is close to half-wavelength, and thus we can use (\ref{L_strips}).

Although in principle it is possible to write exact analytic formulas for all the vector components of ${\overline{e}}_t^\mathrm{f}$, ${\overline{e}}_r^\mathrm{f}$, ${\overline{e}}_t^\mathrm{b}$ and then design the system analytically to reach the goals 1--4, it is not practical since the formulas become very cumbersome. Instead, we use a numerical optimization approach. For optimization, the central frequency was chosen to be equal to $100 \, \mathrm{GHz}$, so $\omega_0= 2\pi \cdot 100\cdot10^9 \, \mathrm{Rad/s}$. The optimization variables are as follows: 1) the period of the strips $d$, the same for both grids; 2) the modulation frequency $\omega_{m}$, the same for both grids; 3) the width of the strips of grid 1; 4) the width of the strips of grid 2 (these widths affect the equivalent inductance of each grid); 5) the modulation parameter $\xi_{1}$ (see Eq. (\ref{xi})) for grid 1; 5) the modulation parameter $\xi_{2}$ for grid 2; 6) the phase delay for the modulation function of grid 2; 7) the thickness of the air gap 1; 8) the thickness of the air gap 2; 9) the thickness of the dielectric slab. The thickness of the dielectric substrate of each grid was fixed to $0.5$~mm which is usual for wafers.

\subsection{MATLAB optimization}\label{matlab-optimization}

A MATLAB program was created to reach the optimization goals using global optimization. It is based on the code in Supplementary Materials for the work \cite{ref9}. This program optimizes the parameters of the structure to reach  goals 1-4. The global optimization algorithm takes an arbitrary initial point in the bounded area in the space of variable parameters and then moves the point to minimize the objective function, defined as follows:
\begin{equation} 
F = \sqrt{\left| {\overline{e}}_r^\mathrm{f}\left( \omega_0 \right) \right|^2 + \left( \left| {\overline{e}}_t^\mathrm{f} \left( \omega_0 \right) \right| - 1 \right)^2 + \left| {\overline{e}}_t^\mathrm{b}\left( \omega_0 \right) \right|^2},
\end{equation} 
where the use of $e(\omega_0)$ means that the $(N + 1)$-th component of the corresponding vector, which represents the main harmonic, is taken. This compound objective function was chosen to balance the optimization of all three parameters and to avoid over-optimizing one of the parameters separately.

Each run of the optimization routine produces a new set of parameters, corresponding to some local minimum of the global objective function. Since it is theoretically impossible to prove that the global optimization converged to the global minimum, and not to one of the local minima, we run the optimization routine many times and then choose the best goal function value and the corresponding set of the problem parameters. We also have noticed that the best results are achieved when the modulation frequency is not considered as a free optimization parameter but instead fixed to some arbitrary reasonable value prior to the optimization run. From the practical point of view, it is preferable to have the  modulation frequency $\omega_{\mathrm m} \ll \omega_0$. Lowering $\omega_{\mathrm m}$ makes practical implementations of the system easier, but for very low modulation frequencies, the required length of the time-modulated system starts to increase, so it is necessary to keep the balance between those two secondary optimization goals. 

Among 100 consecutive runs, the following set of the parameters was chosen as the best one: $\omega_{\mathrm m}=0.235 \, \omega_0$, the air gap 1 thickness $h_\mathrm{A}^{(1)}= 0.38 \lambda_0 = 1.141 \, \mathrm{mm}$, the air gap 2 thickness $h_\mathrm{A}^{(2)} = 0.514\lambda_0=1.543 \, \mathrm{mm}$, $\xi_1 = 0.532$, $\xi_2 = 0.497$, the  dielectric slab thickness $h_\mathrm{D}^{(2)} =0.473\lambda_0=1.421 \, \mathrm{mm}$, the phase delay between the two grids $\varphi= 1.681 \, \mathrm{rad}$, the grid period (the same for both grids) $d = 0.182\lambda_0=0.546 \, \mathrm{mm}$, the strip width for grid 1 $w_1 = 0.041\lambda_0 = 0.123 \, \mathrm{mm}$, the strip width for grid 2 $w_2 = 0.048\lambda_0 = 0.143 \, \mathrm{mm}$ the strip thickness (the same for both grids) $h_{\mathrm{S}}=0.01 \, \mathrm{mm}$, where $\lambda_0 = 3 \, \mathrm{mm}$ is the free-space wavelength at the central frequency. The period of the gaps along the $X$-axis was set to 0.1~mm. This set of parameters gives the following isolator performance from the analytical model at the central frequency $\omega_0= 2\pi \,100\cdot 10^9 \, \mathrm{Rad/s}$: forward transmission is equal to $-2.1 \,\mathrm{dB}$ (the insertion loss is equal to $2.1 \,\mathrm{dB}$); backward transmission is equal to $-21 \, \mathrm{dB}$ (the isolation is equal to $21 \, \mathrm{dB}$); forward reflection is equal to $-22 \, \mathrm{dB}$. The total length of the device for this set of parameters is about 5.1~mm, which is $\approx 1.7 \, \lambda_0$, where $\lambda_0$ is the free-space wavelength. We can write the system performance in matrix form in dB scale:
\begin{equation}
\begin{array}{cl}
    |\mathrm{\mathbf{S}}|(\mathrm{dB}) & \equiv 20\left( \begin{array}{cc}
     \log_{10}({|S_{11}|})    &  \log_{10}({|S_{12}|}) \\
     \log_{10}({|S_{21}|})    &  \log_{10}({|S_{22}|})
    \end{array} \right) \\ \\
    & = \left( \begin{array}{cc}
     -22    &  -21 \\
     -2.1    &  -21
    \end{array} \right).
    \end{array}
\end{equation}

The amplitudes of the harmonics are presented in Fig.~\ref{fig:Fig2}b. Most of the incident energy for the backward propagation and forward reflection is converted to  other harmonics $\omega_0\pm n\omega_{\mathrm m}, \, n=\pm1, \pm2, \ldots \pm N$, where it can be easily filtered out by a bandpass filter with the central frequency of $\omega_0 = 2\pi \cdot 100\cdot 10^9 \, \mathrm{Rad/s}$. For the forward propagation, most of the energy is preserved in the central harmonic $\omega_0$. Together, these properties characterize  nonreciprocal isolation, where only one propagation direction of energy at the fundamental frequency $\omega_0$ is allowed. 

\section{Full-wave numerical simulations}

\subsection{Free-space simulations}

Full-wave simulations of a multi-layered structure with the geometry and modulation parameters obtained in Section~\ref{matlab-optimization} were performed with the help of CST Microwave Studio 2025. A simulation domain enclosing one period of a strip grid in both directions on the grid plane was considered. Together with appropriate boundary conditions this allows simulating an infinite plane grid for a fixed polarization of the incident waves. For the electric field parallel to the $X$ axis, Perfect Magnetic Conductor (PMC) boundary conditions on the boundaries orthogonal to the $Y$ axis and Perfect Electric Conductor (PEC) boundary conditions on the boundaries orthogonal to the $X$ axis together efficiently mimic a structure infinite in the $X$ and $Y$ directions. A plane wave at the frequency $f_{0} = 100\ \mathrm{GHz}$ is impinging on the structure in the direction of the $Z$ axis. Open boundaries in both positive and negative directions of the $Z$ axis were simulated with the help of two waveguide ports. The structure representing a single cell of an infinite grid consists of a copper strip with a gap. The gaps are filled with an artificial material with a time-modulated conductivity, which mimics externally driven nanoplasma discharges.

\begin{figure*}
\begin{center}
\includegraphics[width=0.99\textwidth]{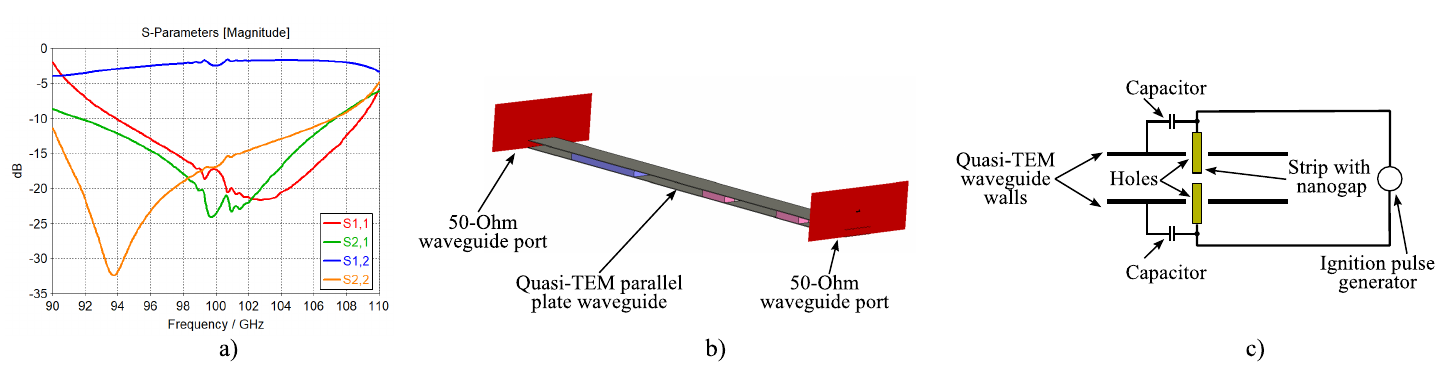}
\caption{\label{Fig:fig3} a) 
Simulated magnitudes of S-parameters of a time-modulated isolator; b) Schematic of the parallel-plate  waveguide model in CST Microwave Studio; c) Schematic of a biasing voltage isolation in a parallel-plate waveguide.}
\end{center}
\end{figure*}

The simulated values of S-parameters of the system are presented in (\ref{S_simulated}). The system shows a strong non-reciprocal behavior, with insertion loss $2.5 \, \mathrm{dB}$ and isolation of $23.5 \, \mathrm{dB}$, which is in good agreement with the values predicted by the theory. Reflection, however, is higher than predicted, still being in satisfactory agreement with the prediction.  The bandwidth of isolation, measured at the level of  $-20 \,\mathrm{dB}$, is about 3.8\% of the central frequency. The plots of all the four components of the S-matrix as functions of the  frequency in the vicinity of $100 \, \mathrm{GHz}$ are presented on Fig.~\ref{Fig:fig3}a. 
\begin{equation}\label{S_simulated}
    |\mathrm{\mathbf{S}}|(\mathrm{dB})  = \left( \begin{array}{cc}
     -17    &  -23.5 \\
     -2.5    &  -16.9
    \end{array} \right).    
\end{equation}

\subsection{Waveguide: rectangular vs parallel-plate}\label{waveguide-rectangular-vs-parallel-plate}

The theoretical model proposed above explicitly assumes that the metasurfaces are infinite in the $X$ and $Y$-directions. In a rectangular waveguide, this model is no longer valid since it does not support TEM modes. In the simulations above, we mimic infinite structures by an appropriate combination of PEC and PMC boundaries. Since PMC boundaries are hard to implement in real-life experiments, we switch to a waveguide that supports TEM modes. One of the possible implementations of a TEM waveguide is an open-boundaries parallel-plate waveguide~\cite{met4}, which is shown in Fig.~\ref{Fig:fig3}c. When the distance between the plates $h \ll \lambda_0$, where $\lambda_0$ is the free-space wavelength, this  waveguide supports a TEM mode, which has the internal field structure similar to that of a plane wave in free space. This method for measuring the properties of metasurfaces was considered in, for instance, \cite{met2}. In our simulation, the structure is excited by two waveguide ports. Although the field structure between the two plates is similar to the free-space field, the electromagnetic field is not fully confined inside the wavegide cross-section. This effect, called fringing field effect, affects the total performance of the TEM waveguide as an equivalent of free-space propagation. The impact of the fringing fields can be reduced by increasing the width of plates in the transversal direction. In our case, the width of the waveguide must be proportional to the grid period $d$. We have conducted simulations for three different plates widths: $d$, $2d$, and $3d$. Simulated $S$-matrices for each case are presented in Eq.~\eqref{27}, where $|\mathrm{\mathbf{S}}^{(1)}|$ stands for 1 period of the grid, $|\mathrm{\mathbf{S}}^{(2)}|$ for two periods and $|\mathrm{\mathbf{S}}^{(3)}|$ for three periods. As we can see, one period of the grid is not sufficient to obtain the performance equal to the free-space performance, due to the fringing field effect. Two- and three-period model show the performance much closer to the performance of an infinite grid.

\begin{equation}
\begin{array}{cc}
|\mathrm{\mathbf{S}}^{(1)}|(\mathrm{dB})  = \left( \begin{array}{cc}
     -15.7   &  -15.3 \\
     -3.2   &  -7.1
    \end{array} \right),
     \\ \\
     |\mathrm{\mathbf{S}}^{(2)}|(\mathrm{dB})  = \left( \begin{array}{cc}
     -15   &  -22.5 \\
     -3.2   &  -10
    \end{array} \right), 
    \\ \\
    |\mathrm{\mathbf{S}}^{(3)}|(\mathrm{dB})  = \left( \begin{array}{cc}
     -19.3   &  -28 \\
     -2.7   &  -12.9
    \end{array} \right),    
\end{array}
\label{27}
\end{equation}

\subsection{Nanoplasma gaps biasing circuit}\label{A biasing voltage isolaton from the waveguide walls}

In the model presented in the previous section, modulated strips affect the waveguide mode propagation only if they are electrically connected to the waveguide walls, otherwise the wave does not interact with the strip in either state of the switches. However, this means that we should apply the DC voltage that switches the nanoplasma discharge on and off directly to the waveguide walls, which is highly undesirable. To solve this problem, we propose to use capacitive coupling of the strips to the waveguide walls, which ensures the interaction of the mode with the strips, but isolates the waveguide walls from the modulating voltage. The schematic is shown in Fig.~\ref{Fig:fig3}d. The value of the coupling capacitance was obtained by a numerical optimization. Full-wave simulations show that this method of isolation does not significantly affect the performance of the multilayered time-modulated isolator.

\section{Conclusion}

In this article, we have developed an analytical framework based on the time-Floquet method for design of nonreciprocal traveling-wave devices based on time-modulated metasurfaces built with two-state time-modulated elements -- nanoplasma-based switches. Such metasurfaces exhibit a step-wise temporal modulation, which makes this  solution and equivalent circuit distinct from previous works. We consider a wire or strip grid with nanogaps as the metasurface for our model. The nanoplasma is generated by an external bias voltage inside the gaps in metallic wires. Nanoplasma discharge gaps, which were shown experimentally to have a picosecond-scale swithcing time and very small capacitance, allow us to avoid common disadvantages of widely used semiconductor switches and varactors, thus enabling much higher modulation speeds. An example of a double-metasurface isolator with the central frequency of $100 \, \mathrm{GHz}$ was developed and studied both analytically and numerically, with the help of full-wave simulations. A potential realization in a TEM parallel-plate waveguide was also simulated numerically. The results of the full-wave simulations show a very good agreement with the theoretically predicted results.

\section*{Acknowledgments}
This work was supported in part by the Pathfinder Open 2022 program of the European Innovation Council, project number 101099313 (PULSE project), the Research Council of Finland within the RCF-DoD Future Information Architecture for IoT initiative
(grant no. 365679), and the Fundamental Research Funds for the Central Universities, China (project no. 3072024WD2603).

\bibliography{main}

\end{document}